\documentclass[epj,final]{svjour}

\usepackage{latexsym}
\usepackage{url}
\usepackage{amsfonts}
\usepackage{amsmath, amssymb}
\RequirePackage{graphicx}

\begin{document}

\title{On the Galactic CMB apex}

\author{V.G. Gurzadyan\inst{1,2} \and A.L.Kashin\inst{1} A.A.~Kocharyan\inst{3}\and A. Stepanian\inst{1}}

\institute{Center for Cosmology and Astrophysics, Alikhanian National Laboratory and Yerevan State University, Yerevan, Armenia \and SIA, Sapienza Universita di Roma, Rome, Italy \and 
School of Physics and Astronomy, Monash University, Clayton, Australia  
}
                  

\abstract{The hierarchy of motions that we are participating is well known, from the Earth's motion around the Sun and Sun's motion in the Milky Way, up to the Local Group's motion within the Virgo Supercluster of galaxies. The dipole anisotropy of the Cosmic Microwave Background (CMB) enables to define Sun's motion with respect to the CMB "absolute" frame. We now present evidence for CMB Galactic dipole signal due to the Sun's motion in the Galaxy, by means of Planck 2018 data and Gaia Early Data Release 3. The signal is weak and frequency depended, the strongest is at 30 GHz, up to $7.6 \sigma$ confidence. The amplitude of the signal interpreted as Doppler caused, corresponds to velocity $v \approx 225.5 \pm 16.2 \, km\, sec^{-1}$, in agreement with Solar system's velocity with respect to the Galactic center. While the revealing of precise coordinates of the apex will need further refined analysis at various bands, the detected weak signal can indicate the appearance of a new cosmic scaling in CMB, thus opening a link to a bunch of physical effects.}


\maketitle

\section{Introduction}

William Herschel \cite{H} was the first to consider the Sun's motion and its direction within the surrounding stars. Since then, the long span of investigations culminates at the recent {\it Gaia Early Data Release 3}, having revealed the Solar system acceleration  in direction $l=358.9^{\circ} \pm 4.1^{\circ}, b=-3.3^{\circ} \pm 4.6^{\circ}$, with respect to 1.6 mln remote compact extragalactic sources \cite{Gaia} (also includes an historical account, with vast references, of early and modern studies of the Solar system motion). This acceleration is composed mainly by the Sun's motion in the Galaxy, although  other, more tiny, contributions are not excluded. 

Today, the Solar apex and, the local standard of rest (LSR) and the associated hierarchy of motions, i.e. Sun's motion with respect to the Galactic center, within the Local Group, the Local Supercluster, are revealed securely. A unique calibration to this problem is provided by the Cosmic Microwave Background (CMB) and its temperature dipole (l=1) anisotropy \cite{Pl}, defining Sun's motion with respect the CMB "rest" frame. Among various aspects of the importance of the mentioned hierarchy of motions is the recently arisen Hubble constant tension,  as tension between the late and early Universe, see \cite{VTR,R} and references therein.   This is a principal issue, since as shown in \cite{GS2,GS3}, the Hubble tension can be solved considering the cosmological constant in the weak-field limit of General Relativity and being responsible for the recession of galaxies in the local Universe -- the local Hubble flow -- differing from the cosmological expansion of the Universe i.e. the Global flow by its nature.

In this study, we use the Planck 2018 data \cite{Pl} to analyse the possible contribution of the Gaia revealed Solar system's motion in the CMB maps; the Planck determined CMB dipole at 100 GHz is of amplitude $3362.48 \pm 0.10 \mu K$ and apex $l=264.022^{\circ} \pm 0.006,\, b=48.253^{\circ} \pm 0.003$; the blackbody temperature of CMB is $T_{CMB} = 2.72548 K \pm 0.57 mK$ \cite{Fi}.

We present the detection of a trace of CMB dipole anisotropy towards the Gaia's above mentioned apex of the Solar system. The amplitude of the anisotropy agrees with the Sun's peculiar velocity with respect to the Galactic center. 

\section{Planck data analysis} 

The amplitude of the dipole $l=1$ anisotropy, as of the most strong signal in CMB sky and interpreted as our motion with respect to CMB frame, corresponds to velocity $v= 369.82 \pm 0.11\, km\, sec^{-1}$ \cite{Pl}. There is certain frequency dependence both of the amplitude and the apex coordinates of the dipole anisotropy (Table 6 in \cite{Pl}), determined by calibration and other effects. The signal we are looking for here, has to be of weaker amplitude than that of the CMB dipole and, especially, less sounding and frequency depended due to the location of the Gaia's apex close to the Galactic equator and hence due to various contributions to the signal.

For our analysis we used the data of Planck's final release (2018) \cite{Pl} of 30, 70, 100, 143 GHz frequencies. Fig.1 shows the CMB temperature difference of regions surrounding the Gaia apex $l=358.9^{\circ} \pm 4.1^{\circ}, b=-3.3^{\circ} \pm 4.6^{\circ}$ and antapex, averaged within circles of $(0.5^{\circ}, 1^{\circ}, 2^{\circ}, 3^{\circ})$ radii. In Fig.2 we present the similar analysis of the temperature difference but averaged for a sequence of 72 regions composed by 5 degree shift from the Gaia apex and antapex, at the same Galactic latitude. In Figs.1,2 we show the same analysis results also for SMICA maps (for details on SMICA see \cite{Pl} and references therein), which are produced by combining the input channels and using a mask for the Galactic disk and hence are useful for search of details in the cosmological signal (e.g. \cite{GSt}). The drastic difference of the SMICA results from those of other frequencies leaves no doubt in the Galactic nature of the latter.     

Fig.3 exhibits the values of the apex-antapex temperature difference for the 360 regions of radius $4^{\circ}$ at $4^{\circ}$ shift and their antapexes, for 30 GHz. Fig.3 thus enables to reveal the variation of the apex-antapex temperature difference over the sky at those Galactic latitudes. For comparison, Fig.4 shows the same as in Fig.3 but for $2^{\circ}$ at $5^{\circ}$ shift. The data in both Figs.3,4 are fitted with 6th order polynomial. In Fig.5 we give the temperature difference scan for 360 regions of  $4^{\circ}$ radii, for 3 sky periods, with standard errors and 13th order polynomial fit.
  
\begin{figure}[h]
\caption{CMB temperature difference -- of Gaia's apex and antapex -- for Planck's bands, 30, 70, 100, 143 GHz and for SMICA, averaged within circles of radii $(0.5^{\circ}, 1^{\circ}, 2^{\circ}$); the standard error bars are indicated.}
\centering
\includegraphics[width=9cm]{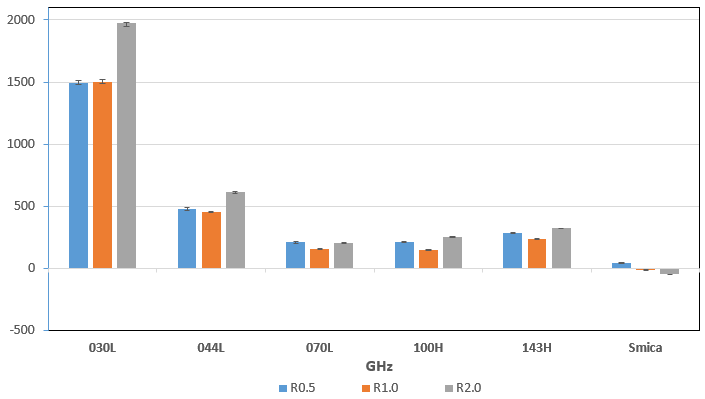}
\end{figure}

\begin{figure}[h]
\caption{The same as in Fig.1, but temperature differences for 72 circular regions, composed at 5$^{\circ}$ shift, at the same Galactic latitude as of Gaia's apex.}
\centering
\includegraphics[width=9cm]{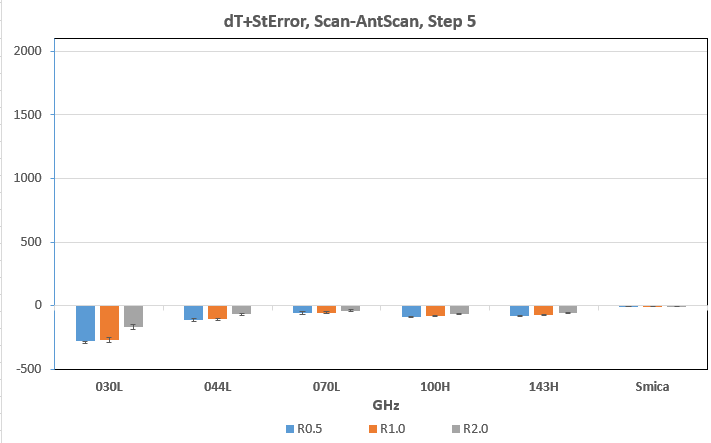}
\end{figure}

\begin{figure}[h]
\caption{The apex-antapex CMB temperature difference as in Fig.2 but for the 360 circular regions of 4$^{\circ}$ radii, at 1$^{\circ}$ shift, for 30 GHz.}
\centering
\includegraphics[width=9cm]{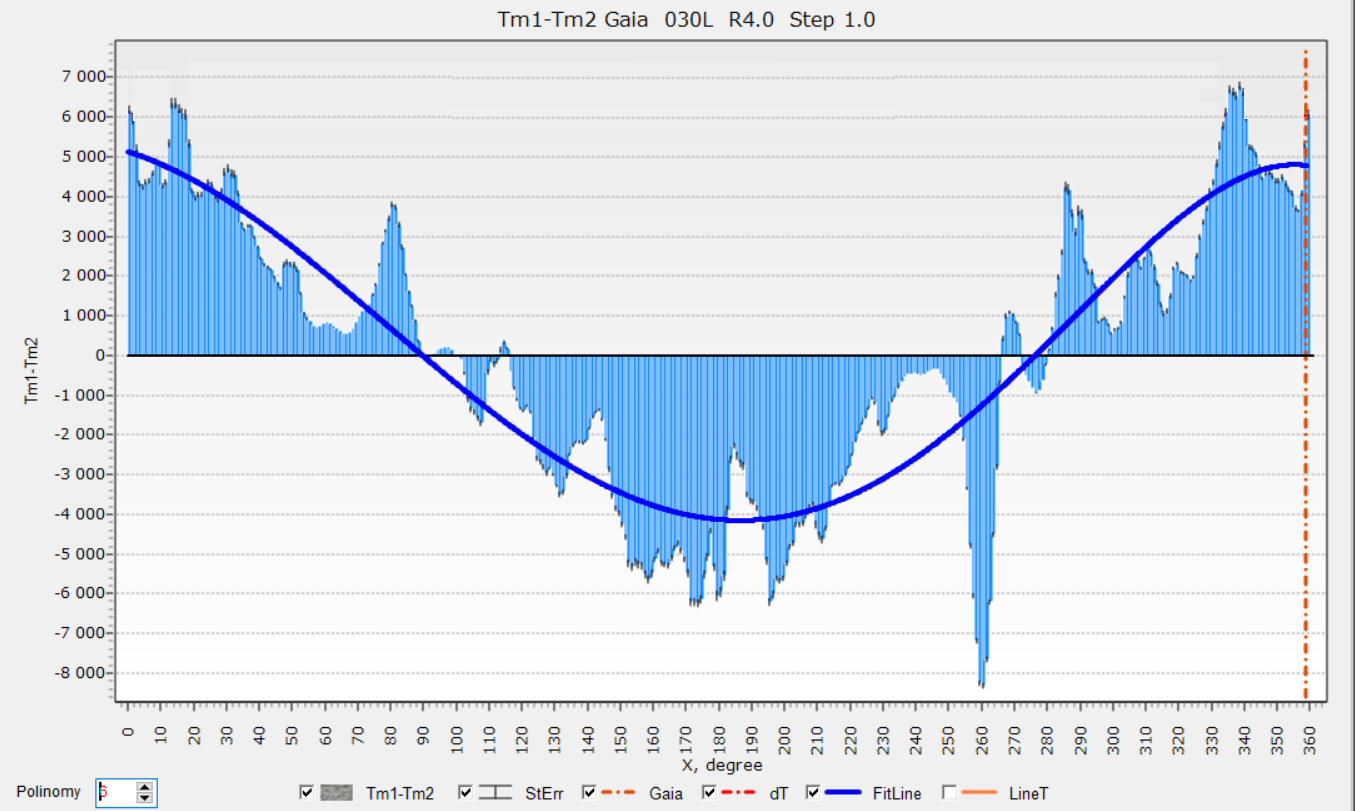}
\end{figure}

\begin{figure}[h]
\caption{The same as in Fig.3, but for 2$^{\circ}$ radii 72 circular regions, at 5$^{\circ}$ shift, for 30 GHz; the bars correspond to standard error.}
\centering
\includegraphics[width=9cm]{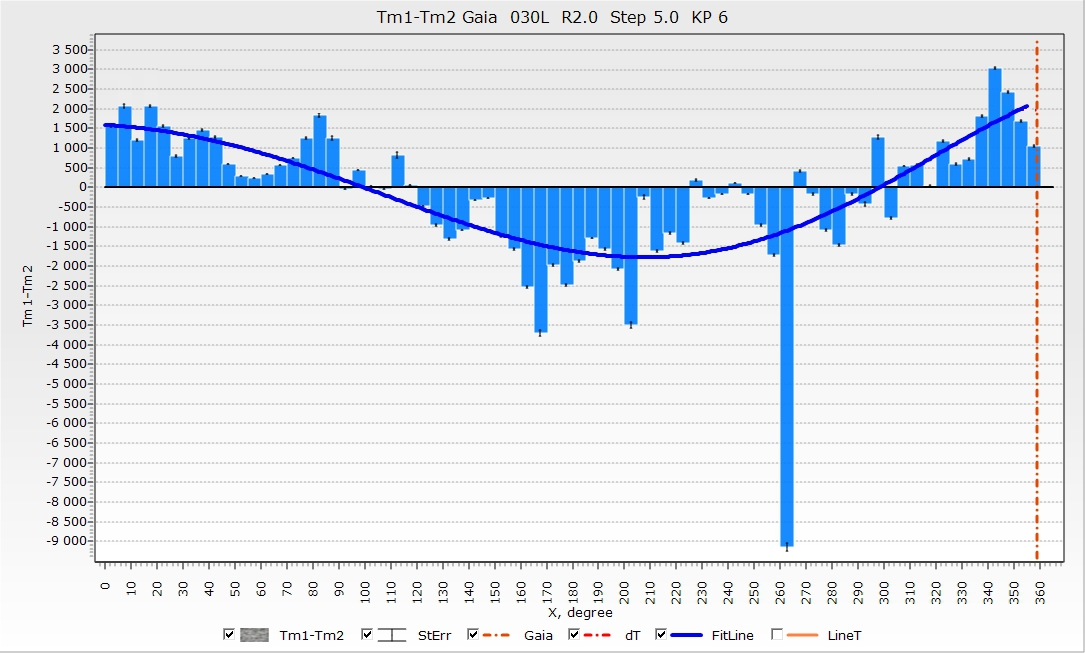}
\end{figure}

\begin{figure}[h]
\caption{The same as in Figs.3 and 4, but for 4$^{\circ}$ radii 360 circular regions, formed at 1$^{\circ}$ shift, for three 360$^{\circ}$ periods, with standard errors.}
\centering
\includegraphics[width=9cm]{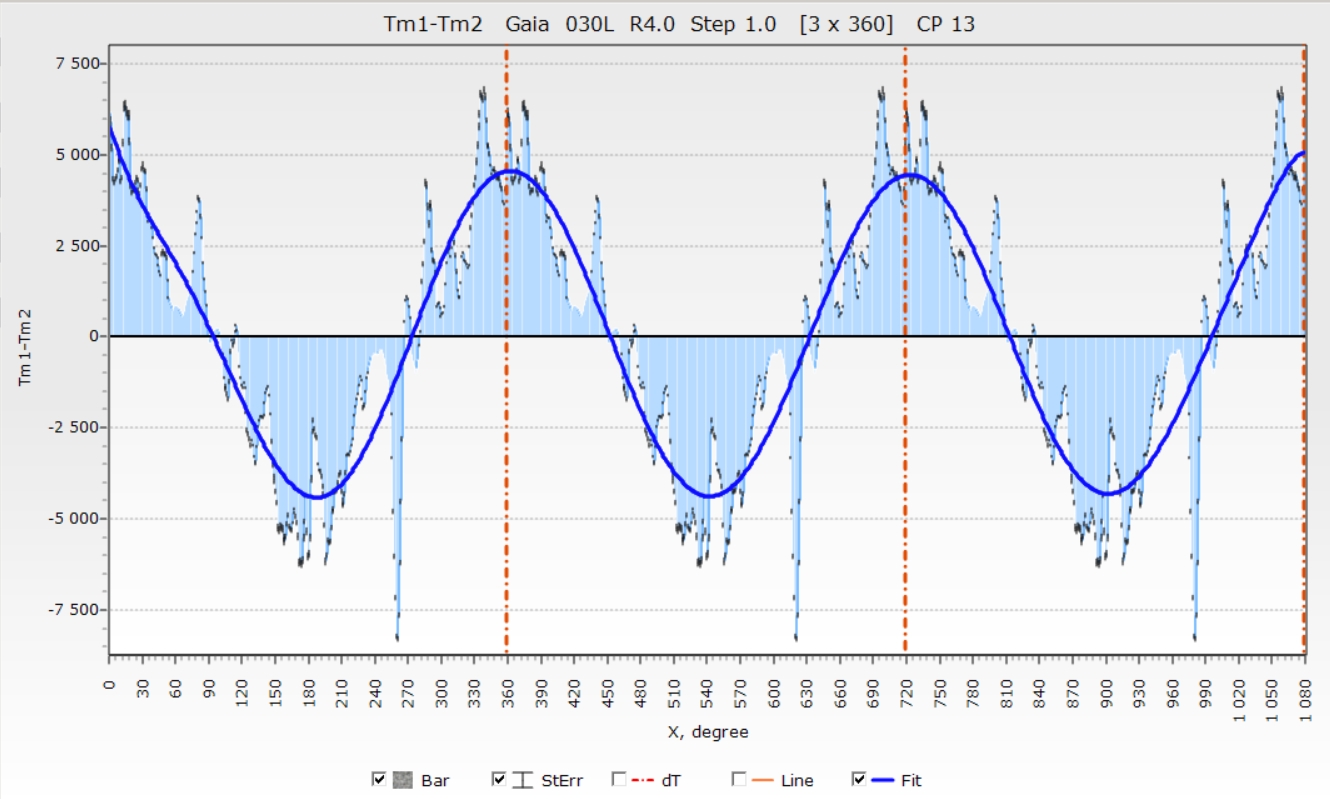}
\end{figure}



\section{Conclusions}

We studied the CMB temperature anisotropy maps of Planck 2018 in the vicinity of the apex for the Solar system acceleration provided by Gaia Early Data Release 3 \cite{Gaia}. The CMB temperature difference for Gaia apex and antapex, as well as for 360 and 72 regions of the same Galactic latitude and shifted on 1 and 5$^{\circ}$, respectively, are obtained. 

The analysis does show a weak signal, the strongest at 30 GHz, when for two regions of radius 0.5$^{\circ}$ at Gaia apex-antapex directions the temperature difference yields $1496.5 \mu K$ with	$\sigma=196.8 \mu K$, and 1$^{\circ}$ it is $1504.0 \mu K$ and $420.8 \mu K$ i.e. $7.6\sigma$ and $3.57\sigma$, respectively. The fit of apex-antapex temperature difference for 72 regions (Fig.4), for example, yields for the apex access temperature the maximum amplitude $\approx 2046.7 \pm 14.7 \mu K$ at $l=356 \pm 0.5^{\circ}$ ($b$ is that of Gaia's apex in the used procedure).

Thus, the weak dipole signal in CMB Planck maps associated to Gaia apex region, depends on the frequencies as expected, given the sky location of the Gaia apex. For a moving observer the dipole term is the first one in the right hand side of radiation temperature distortion formula 
\begin{equation}
\delta T/T_{CMB}= (v/c) \cos \theta + (v/c)^2 \cos 2\theta + o(v/c)^3, \,\, v<<c.
\end{equation}
Using the above mentioned amplitude for the temperature signal and attributing it via this formula to a Doppler velocity, we have $v \approx 225.5 \pm 16.2 \, km\, sec^{-1}$ (cf.\cite{Ma}).  

The frequency dependence of the measured signal, i.e. of its amplitude and apex (including difference from Gaia's apex), certainly needs refined analysis of variety of physical effects \cite{Pl1}, including those associated to the optical depth $\tau(\nu,r)$
\begin{equation}
\delta T/T_{CMB}= (v/c) \int \tau(\nu,r) dr.
\end{equation}

Our goal here is to present the evidence for Galactic CMB dipole signal, particularly, in view of its importance for the hierarchy of motions that we are participating, from the Solar motion in the Galaxy, up to  the Local Group's in Virgo Supercluster. The latter, as mentioned, can be linked to a basic cosmological issue, the Hubble tension \cite{VTR,R} and the local Hubble flow (e.g. \cite{RG,GK,GS3,Ch} and references therein).

{\it Acknowledgments.} The use of the {\it Planck} data in the Legacy Archive for
Microwave Background Data Analysis (LAMBDA)  and of HEALPix package \cite{Go}
is acknowledged.

\end{document}